\documentstyle[12pt]{article}

\begin{document}

\title{Using the radiative decay $b\to s\gamma$ to bound the
chromomagnetic dipole moment of the top quark}
\author{R. Mart\'{\i}nez and J-Alexis  Rodr\'\i{}guez \\
Universidad Nacional,
Depto. de F\'\i{}sica \\
Bogot\'a, Colombia}
\date{}

\maketitle

\begin{abstract}
Using the CLEO data for $b\to s\gamma$, we constrain the chromomagnetic 
dipole moment of the top quark within the context of a decoupling effective
Lagrangian approach. Our results are in agreement with the results
obtained for the $e^+e^-$ and hadron colliders.
\end{abstract}

\newpage

The Standard Model (SM) describes successfully all 
experimental data related with the strong and electroweak interactions.
In particular, the electroweak radiative corrections at one loop
level \cite{berna} agree consistenly with the top quark mass measured by 
the CDF and D0 collaborations \cite{CDF}. However, the hierarchy mass
problem
suggests that new physics effects may lay beyond the SM. 

Since the top quark mass is so heavy, it is expected that its physics 
may be different from the lighter fermions and that the top quark
might couple quite strongly to the electroweak symmetry breaking sector. 
This suggests
that the Higgs sector of the SM is just an effective theory and that
the physics beyond the SM may be manifested through an effective Lagrangian 
involving the top quark \cite{peccei}.

The framework of effective theories, as a mean to parametrize physics beyond the
SM in a model independent way, has been used recently. Two cases have been 
consider in the literature, the decoupling case, which includes the Higgs boson
\cite{wyler}, and
the non-decoupling case, where there is no Higgs boson. We shall consider only
the first case, in which the SM is a low-energy limit of a renormalizable
theory.
In this approach, the effective theory parametrizes the effects at
low energy of the full renormalizable theory by means 
of high order dimensional non-renormalizable operators written in 
terms of the SM fields.

In the present work we are interested in studying possible deviations
from the SM on the decay $b\to s\gamma$ within the context of the effective
Lagrangian approach. Several authors have been used the CLEO 
results on radiative 
$B$ decays to set bounds on the anomalous coupling of the
t-quark \cite{han,nos}. While in the first case \cite{han} they used a 
parametrization which is only
$SU(3)_C$ or $U(1)_Q$ 
gauge invariant to get bounds on the anomalous magnetic dipole moments, in
the second ref.
\cite{nos} they used one which is SM gauge invariant. In the present brief
report,
we will use dimension-six operator
which are full strong and electroweak gauge invariant and contribute to 
$b\to s\gamma$ in order to bound the chromomagnetic dipole moment of
the top quark. Since
this approach includes an effective Lagrangian respecting 
the full SM symmetry, we should expected some differences in the calculation 
of the $b\to s\gamma$ rate obtained from the mere use of $SU(3)_C$ gauge
invariant
effective operators.

If an anomalous top quark coupling exists, it will modify the SM 
expectations for the
top quark production and decay processes at hadron and $e^+e^-$ colliders
\cite{cpy}. In ref. \cite{rizo1} it 
was found that $-0.027 < \delta\kappa_g^t < 0.026$ for the process 
$e^+e^-\to t\bar{t}g$ at a $500$ GeV NLC with an integrated luminosity of
$100$ fb$^{-1}$ and a gluon spectrum above $E_g^{min}=25$ GeV. On the other
hand, using the $\bar{t}t$ invariant mass and the $p_t$ distribution of
the top
quark at the LHC, the bound
$\mid\delta\kappa_g^t\mid < 0.05$ was obtained with an $100$ fb$^{-1}$
integrated luminosity \cite{rizo2}. Similarly, the CDF data on the branching 
decay to $b$
was used to get the limit $BR(t\to cg)<0.45$. This bound in turn gives the limit 
$\mid\kappa_g^{ct}\mid < 0.9$ for the neutral current flavour changing 
transition moment \cite{han}. It is important thus to consider different
approaches to study the top quark
anomalous couplings that may allow us to understand the physics beyond the SM. The
prevailing differences  of the SM results with respect to the 
experimental data for the partial width ratios
$\Gamma(Z\to b\bar{b}(c\bar{c}))/\Gamma(Z\to hadrons)$ suggest also to 
consider physics
beyond the SM.

The effective Lagrangian involving the anomalous top quark couplings can be written
as 
\begin{equation}
{\cal L}={\cal L}^{SM} + \Delta{\cal L}^{eff} \; ,
\end{equation}
where ${\cal L}^{SM}$ is the SM Lagrangian and $\Delta{\cal L}^{eff}$ includes
the anomalous quark couplings. We consider the following dimension-six, CP-conserving
operators, which are $SU(3)_C\otimes S(2)_L\otimes U(1)_Y$ gauge invariant
\begin{equation}
{\cal O}^{ab}_{uG}=\bar{Q}_L^a \sigma_{\mu\nu} G^{\mu\nu\; i}
\frac{\lambda^i}{2}\tilde{\phi} u_R^b \; ,
\label{six}
\end{equation}
where $G^{\mu\nu\; i}$ is the gluon field strength tensor and $a, b$ are
the
family indices. The above operator gives rise to the anomalous $t\bar{t}g$ 
vertex and its respective 
unknown coefficient $\epsilon_{ab}^{uG}$ is related with the anomalous
chromomagnetic 
moment of the top quark by
\begin{equation}
\delta\kappa^t_g =\sqrt{2}\frac{g}{g_s}\frac{m_t}{M_W}\epsilon_{uG}^{33} \; .
\end{equation}

The effective Hamiltonian used to compute the $b\to s$ transition is given
by \cite{buras}
\begin{equation}
H_{eff}=-\frac{4 G_F}{\sqrt{2}} V_{ts}^{*} V_{tb} \sum_{i=1}^{8} c_i(\mu)
{\cal O}_i(\mu) \; ,
\end{equation}
where $\mu$ is the energy scale at which $H_{eff}$ is applied. For
$i=1 - 6$, ${\cal O}_i(\mu)$ correspond to four-quark operators, ${\cal
O}_7(\mu)$
is the electromagnetic dipole moment and ${\cal O}_8(\mu)$ is the chromomagnetic
dipole operator. At low energy, $\mu\approx m_b$, the only operator that
contributes
to $b\to s$ transition is ${\cal O}_7(\mu)$ which results from a mixing
among the ${\cal O}_2(M_W)$,
${\cal O}_7(M_W)$ and ${\cal O}_8(M_W)$ operators.

The six-dimensional operator (\ref{six}) contributes to ${\cal O}_8$
by
the diagrams shown in Fig. 1. The contribution of the diagrams $(1a)$, $(1b)$
and $(1c)$ are finite, while the diagram $(1d)$ has a logarithmic divergence. We replace
the pole $1/\epsilon$ by $\ln(\Lambda^2)$, where $\Lambda$ can be 
interpreted as the
scale associated with the new physics. In the approaches used in ref. \cite{han},
the contribution to $\delta\kappa_g^t$ comes from diagrams (1a) and (1b). On
the other hand, the operator (2) gives additional contribution given by
diagrams (1c) and (1d). 

The total contribution of the effective operator (\ref{six}) to
the ${\cal O}_8(M_W)$ operator can be written as:
\begin{equation}
c_8(M_W)=c_8(M_W)^{SM}+\delta\kappa_g^t \Delta c_8(M_W) \; ,
\end{equation}
where
\begin{eqnarray}
\Delta c_8(M_W)&=& \frac{1}{4 V_{ts}}
\ln\left(\frac{\Lambda^2}{M_W^2}\right)
+\frac{1}{V_{ts}}\frac{x-x^2+x (2-x)\ln(x)}{8(1-x)^2} \nonumber \\
&+&\frac{3 x - 4 x^2 + x^3 + 2 x\ln(x)}{8 (1-x)^3}  
\end{eqnarray}
and $x=m_t^2/M_W^2$.

In Fig. 2 we display the $b\to s\gamma$ branching fraction as function of
$\delta
\kappa_g^t$ with $m_t=180$ GeV (dashed line). The solid line is the result
of the CLEO
collaboration on the inclusive quark level process, $1\times 10^{-4}
<BR(b\to s\gamma)<4.2\times 10^{-4}$ \cite{cleo}. We find that 
our bound
$\mid \delta\kappa_g^t\mid < 0.066$
is consistent with analysis done in ref. \cite{rizo1,rizo2}
for the process $e^+e^-\to t\bar{t}g$ and the par production of the top 
quark at the LHC.

The using effective Lagrangian approach, with fully SM gauge
invariant operator, we have obtained the bound 
$\mid \delta\kappa_g^t\mid < 0.066$, which is consistent with the
results obtained for the $e^+e^-$ and hadron colliders.

We acknowledge helpfull discussions with M.A. Perez. The present work was
complete during a visit to the Physics Department
of CINVESTAV. It was supported by COLCIENCIAS and CONACYT.

\newpage

\begin{center}
{\bf \large Figure Captions}
\end{center}
\vspace{2cm}

\noindent Figure 1. Feynman diagrams contributing to the $b\to s\gamma$
transition. The heavy dots denote an effective vertex.

\vspace{1cm}

\noindent Figure 2. The branching ratio $BR(b\to s\gamma)$ as a function
of
$\delta\kappa_g^t$ for $m_t=180$ GeV (dashed). The bounds of the CLEO
Collaborations are explicitly indicated.

\end{document}